\documentclass[twocolumn,showpacs,prx,floatfix]{revtex4}

\usepackage{amsmath}
\usepackage{amssymb}
\usepackage{lipsum}
\usepackage{bm}
\usepackage{graphicx}

\newcommand{\ddt}[2]{\frac{\mathrm{d^2}#1}{\mathrm{ d}#2^2}}
\newcommand{\ddx}[2]{\ddt{}x}
\newcommand{\ddy}[2]{\ddt{}y}
\newcommand{\ddz}[2]{\ddt{}z}
\newcommand{\pd}[2]{\frac{\partial #1}{\partial #2}}

\newcommand{\fv}[1]{\left\langle #1 \right\rangle}
\newcommand{\Free}{\mathcal{F}}

\renewcommand{\Im}{\operatorname{Im}}

\begin{document}
\title{Dislocation dynamics and crystal plasticity in the phase field crystal model}

\author{Audun Skaugen\footnote{audun.skaugen@fys.uio.no} and Luiza Angheluta}
\affiliation{Department of Physics, University of Oslo, P.O. Box 1048 Blindern, N-0316 Oslo, Norway}
\author{Jorge Vi\~nals}
\affiliation{School of Physics and Astronomy, University of Minnesota, 116 Church St. SE, Minneapolis, MN 55455, USA}

\begin{abstract}
A phase field model of a crystalline material at the mesoscale is introduced to develop the necessary theoretical framework to study plastic flow due to dislocation motion. We first obtain the elastic stress from the phase field free energy and show that it obeys the stress strain relation of linear elasticity. Dislocations in a two dimensional hexagonal lattice are shown to be composite topological defects in the amplitude expansion of the phase field, with topological charges given by the Burgers vector. This allows us to introduce a formal relation between dislocation velocity and the evolution of the coarse grained envelopes of the phase field. Standard dissipative dynamics of the phase field crystal model is shown to determine the velocity of the dislocations. When the amplitude equation is valid, we derive the Peach-Koehler force on a dislocation, and compute the associated defect mobility. A numerical integration of the phase field crystal equations in two dimensions is used to compute the motion of a dislocation dipole, and good agreement is found with the theoretical predictions.
\end{abstract}
\date{\today}


\maketitle

\section{Introduction}

The description of complex plastic response in crystals at a mesoscale level poses fundamental challenges because of collective effects in dislocation dynamics that give rise to multiple-scale phenomena, such as spatio-temporal dislocation patterning~\cite{Weiss2003,Zaiser2005} and intermittent deformations~\cite{Weiss2015}. Different multiscale models including discrete dislocation models, stochastic models, and cellular automata have been proposed and used to explore various aspects of collective dislocation dynamics~\cite{groma2010statistical,Truskinovsky2011,Mikko2014}. However, these models typically rely on phenomenological input for the dislocation kinetics and mobility, which are important properties and therefore preferably should emerge from theory.

A mesoscale theory is also timely given that defect imaging techniques are beginning to reveal strain and rotation fields created by one or a small number of defects in atomic detail. High Energy Diffraction Microscopy and Bragg Coherent Diffractive Imaging represent the state of the art in imaging at advanced synchrotron facilities \cite{re:rollett17,re:suter17}. The former can provide three dimensional maps of grain orientations with micron resolution, whereas the latter can determine atomic scale displacements with $\leq 30$~nm resolution. Advanced image processing methods allow the determination of the strain field phase around a single defect, clearly evidencing its multivalued nature. Indeed, single dislocations have been successfully imaged and their motion tracked quantitatively just recently \cite{re:yau17}. Experiments also go beyond the determination of strain fields, and determine other quantities sensitive to the topology of the defects. For example, lattice rotation has been imaged and analyzed in nanoindentation experiments \cite{re:sarac16}, or in two dimensional graphene sheets \cite{re:bonilla15}.

Mesoscale models aim at bridging fully atomistic descriptions and macroscopic theory based on continuum mechanics. Along these lines, we mention the so called generalized disclination theory \cite{re:acharya12,re:acharya15}. This theory is a fully resolved nano scale yet continuum dynamical description of dislocations that preserves all topological constraints necessary in the kinematic evolution of the singular fields. Singularities in strains are replaced by topologically equivalent but smooth local fields that allow a full derivation of the governing dynamical equations following the principles of irreversible thermodynamics. The newly introduced fields are similar to a phase field model, except that they are constructed to satisfy all conservation laws, including those of topological origin. On the other hand, the dynamical part of the theory requires constitutive input for both the free energy at the mesoscale, functional of the smooth fields, and mobility relations for their motion.

Conventional phase field models have also become one of the widely used tools in the study of dislocation and grain boundary motion in a wide variety of circumstances. Contrary to the kinematic models, a phenomenological set of dynamical laws for the phase field are introduced, with topological invariants appearing as derived quantities.  There are two different classes of phase field models in the plasticity literature. In one approach, the elementary dislocation is described as an eigenstrain, which is then mapped onto a set of phase fields~\cite{re:wang01,re:koslowski02,re:bulatov06}. If $\mathbf{b}$ is the Burger's vector of the dislocation, and $\mathbf{n}$ the normal to the dislocation line, then the corresponding eigenstrain is defined as
\begin{equation}
u_{ij}^{*} = \frac{b_{i}n_{j}+b_{j}n_{i}}{2a}
\end{equation}
where $a$ is the crystal lattice spacing. The connection to the phase fields $\phi_{\alpha}(\mathbf{x})$, where $\alpha$ label all the slip systems of a particular lattice, is made through the decomposition
\begin{equation}
u_{ij}^{*} = \sum_{\alpha} \epsilon_{ij}^{*\alpha} \phi_{\alpha}(\mathbf{x}). 
\end{equation}
The phase fields are assumed to relax according to purely dissipative dynamics driven by minimization of a phenomenological free energy. This free energy includes a non-convex Ginzburg-Landau type contribution of the same functional form as related studies in fluids~\cite{re:gurtin96}. This contribution is supplemented by an elastic interaction energy that depends only on the incompatibility fields associated with the eigenstrains~ \cite{re:kosevich79,re:nelson81,re:rickman97}, and hence, ultimately, on the phase fields themselves~\cite{re:wang01,re:koslowski02,re:bulatov06}. 

The second approach, which we adopt here, is based on a physical interpretation of the phase field as a temporally coarse-grained representation of the molecular density in the crystalline phase, and is also known as the phase field crystal (PFC) model~\cite{re:elder02,elder2007phase}. The evolution of the phase field is diffusive, and governed by a Swift-Hohenberg like free energy functional, which is minimized by a spatially-modulated equilibrium phase with the periodicity of the crystal lattice. The chosen free energy not only determines the crystal symmetry of the equilibrium phase, but all other thermodynamics quantities and response functions such as its elastic constants~\cite{re:elder02}. As it is generally the case with phenomenological free energies, it is only a function of a few free parameters, and hence the range of physical properties that can be attributed to the macroscopic phase is somewhat limited. Nevertheless, the PFC model has been used in numerous numerical studies including crystal growth, grain boundaries and polycrystalline coarse graining phenomena~\cite{re:eggleston01,re:elder04,re:wu09,re:taha17,Tarp2013}, strained epitaxial films \cite{re:huang08}, fracture propagation~\cite{re:elder02}, plasticity avalanches from dislocation dynamics~\cite{Goldenfeld2010,Tarp2014}, and edge dislocation dynamics~\cite{Berry2006}. It appears to us that this second approach is more natural from a physical point of view in that once the mesoscopic order parameter and the corresponding free energy are introduced, defect variables such as the Burgers vector and slip systems emerge as derived quantities. This seems preferable to introducing Ginzburg-Landau dynamics for slip system amplitudes defined a priori. Also, this second approach can nominally describe highly defected configurations in which a slip system, even in a coarse grained sense, can be difficult to define.

In this paper, we address the important theoretical question as to what extent the PFC model is actually capable of capturing mesoscopic plasticity mediated by dislocation dynamics. Although previous numerical simulations of dislocation dynamics~\cite{Berry2006,Tarp2014} suggest that dislocation motion is controlled by local shear stress, a theoretical derivation from the PFC model is still lacking. Secondly, the diffusive dynamics in the PFC model does not capture the fast relaxation of elastic stresses. To address this question, we consider the PFC model in its amplitude expansion formulation, where we can show that the complex amplitudes are order parameters that support topological defects corresponding to dislocations in the crystal ordered phase. This allows us to accurately define a Burgers vector density field from the topological charges and predict the dislocation velocity directly from the dissipative relaxation of the amplitudes. We show that elastic stresses can be obtained from the PFC free energy functional through standard variational means, and recover known expressions for the linear elastic constants of the medium. Furthermore, we show that the dislocation velocity, at low quenches, follows the Peach-Koehler's force and is given by the Burgers vector and the elastic stress. Our theoretical predictions are consistent with the previous numerical PFC studies of dislocation dynamics~\cite{Berry2006}. However, as recently discussed in Ref.~\cite{Heinonen2014}, the elastic stresses are not at mechanical equilibrium due to the slow, diffusive evolution of the PFC density field, and therefore the overall elasto-plastic response is not captured by the standard PFC model.  

The rest of the paper is structured as follows: in Section II, the phase field crystal model and its elastic equilibrium properties are discussed for the two dimensional case. Here, we also derive the elastic stresses by variational of the free energy functional and express them in terms of the crystal density field. Plastic motion mediated by the dislocation dynamics is treated in Section III, where we use the amplitude expansion and the connection to order parameters supporting topological defects. In Section IV, we verify the theoretical results by direct numerical simulations of the PFC model for a hexagonal lattice with a dislocation dipole. Summary and concluding remarks are presented in Section V. 


\section{Linear elasticity of the phase field model of a crystalline solid}

The phase field crystal model that we employ involves a single scalar field $\psi(\mathbf{x},t)$, function of space $\mathbf{x}$ in two-dimensions (2D) and time $t$, and a phenomenological free energy given by \cite{elder2007phase}
\begin{equation}\label{eq:F}
 \mathcal{F}[\psi] = \int d \mathbf{x} \left[\frac{1}{2} [(\nabla^2+1)\psi]^2 + \frac{r}{2}\psi^2+\frac{1}{4}\psi^4\right],
\end{equation}
where $r$ is a dimensionless parameter. In equilibrium, the free energy functional Eq.~(\ref{eq:F}) is minimized with respect $\psi$, $\mu_{0} = \left( \frac{\delta\mathcal{F}}{\delta \psi} \right)_{0} = 0$ where $\mu$ is the chemical potential and the conjugate variable to $\psi$. When $r > 0$, $\psi = 0$ is the only stable solution, whereas for $r < 0$, equilibrium periodic solutions of unit wavenumber are possible for stripes and hexagonal patterns in 2D \cite{re:elder02}. The crystalline phase with density distribution $n(\mathbf{r})$ is related to the phase field crystal through $\psi(\mathbf r,t) = n(\mathbf r,t)/n_0-1$, where $n(\mathbf r,t) = \sum_i \langle \delta(\mathbf r-\mathbf r_i)\rangle$ is the statistical average number density of the equivalent crystal, and $n_{0}$ its spatially averaged density. 

We focus below on the range of parameters for which a 2D hexagonal lattice is the equilibrium solution~\cite{re:elder02}
\begin{equation}
\label{eq:one_mode_exp}
\psi = \psi_{0} + \sum_{\mathbf q} A_{n}^{(0)} e^{i\mathbf q\cdot \mathbf r}.
\end{equation}
The lowest-order reciprocal lattice wave vectors $\mathbf q_n$ are of unit length in the dimensionless units of Eq.~(\ref{eq:F}), and given in Catersian coordinates by,
\begin{equation}
\mathbf{q}_{1} = \bm{{j}}, \; \mathbf{q}_{2} = \frac{\sqrt{3}}{2} \bm{{i}} - \frac{1}{2} \bm{{j}}, \; \mathbf{q}_{3} = - \frac{\sqrt{3}}{2} \bm{{i}} - \frac{1}{2} \bm{{j}},
\label{eq:reciprocal}
\end{equation}
which satisfy the resonance condition $\sum_{n=1}^3 \mathbf{q}_{n}=0$. The corresponding amplitudes $A_{n}^{(0)}$ are all constant and equal. We next consider a weakly distorted configuration relative to the reference in which both the mean density $\psi_0$ and the amplitudes $A_{n}$ are slowly varying on length scales much larger than the lattice spacing. The distorted phase field can be written as \cite{re:elder10,heinonen2014phase}
\begin{equation}
\psi = \psi_0 + \sum_{\mathbf{q}} A_{\mathbf{q}} e^{i \mathbf{q} \cdot (\mathbf{x} - \mathbf{u})},
\label{eq:distortion}
\end{equation}
where the sum extends over the reciprocal lattice vectors. Note that terms with opposite $\mathbf{q}$ are complex conjugates of one another, so that $\psi$ is real. 

We first examine the change in free energy due to the distortion, and consider $\| \mathbf{u} \|$ small, that is, we do not consider defects. In this case, the mean density $\psi_0$ and the real amplitudes $A_{\mathbf q}$ are uniform and equal to $A_{0}$. The distortion of Eq. (\ref{eq:distortion}) defines the transformation 
\begin{equation}
\mathbf{r} \mapsto \mathbf{r'} = \mathbf{r} + \mathbf{u}(\mathbf{r}) \quad {\rm with} \quad \psi^{\prime}(\mathbf{r'}) = \psi(\mathbf{r}),
\end{equation}
with a free energy functional 
\begin{equation}
  \Free'[\psi'] = \int d^2\mathbf{r}^{\prime} f(\psi', \partial_i' \psi', \partial_{ij}'\psi'),
\end{equation}
where $\partial_{i}^{\prime}$ are partial derivatives with respect to $\mathbf{r}^{\prime}$ and $f(\psi, \partial_i\psi, \partial_{ij}\psi)$ is the integrand in Eq.~(\ref{eq:F}). This free energy can be written in the undeformed coordinate system as,
\begin{equation}
  \Free'[\psi', \mathbf u] = \int d^2\mathbf r \left\| \pd{r'_i}{r_j} \right\|f\left(\psi, \partial_i'\psi, \partial_{ij}'\psi\right). 
\end{equation}
where the Jacobi determinant of the transformation from 
$\mathbf{r}^{\prime}$ to $\mathbf{r}$ is 
\begin{equation}
  \left\| \pd{r'_i}{r_j}\right \| = 
\begin{vmatrix}
  1 + \partial_x u_x & \partial_x u_y \\
  \partial_y u_x & 1 + \partial_y u_y
\end{vmatrix} = 1 + \nabla\cdot \mathbf u + \mathcal{O}(|\nabla u|^2). 
\end{equation}
The derivative terms in the free energy can be transformed by using the expansions
\begin{align}
  \partial_i' &= \partial_i - (\partial_i u_j)\partial_j + \mathcal{O}\left(|\nabla  u|^2\right), \nonumber\\
  \partial_{ij}'\psi &= \partial_{ij}\psi - \partial_i\left[ (\partial_ju_k)\partial_k\psi \right] - (\partial_iu_k)\partial_{kj}\psi + \mathcal{O}\left(|\nabla u|^2\right).
\end{align}
The free energy change $\Delta \Free[\psi, \mathbf{u}] = \Free'[\psi',\mathbf{u}] -\Free[\psi]$ associated with the distortion is 
\begin{widetext}
\begin{align}
\Delta \Free [\psi, \mathbf{u}] = -\int d^2 \mathbf r \left[ \pd{f}{(\partial_i\psi)}(\partial_iu_j)\partial_j\psi + \pd{f}{(\partial_{ij}\psi)}\left\{\partial_i\left[ (\partial_ju_k)\partial_k\psi \right] + (\partial_iu_k)\partial_{kj}\psi \right\} + \left( \nabla \cdot \mathbf{u}\right) f \right] + \mathcal{O}\left(|\nabla u|^2\right).
\end{align}
\end{widetext}
The second term in the r.h.s. can be transformed to a total divergence term and one proportional to the strain,
\begin{widetext}
\begin{equation}
  \pd{f}{(\partial_{ij}\psi)}\partial_i\left[ (\partial_ju_k)\partial_k\psi \right] 
  = \partial_i\left[ \pd{f}{(\partial_{ij}\psi)}(\partial_ju_k)\partial_k\psi \right] - \left( \partial_i \pd{f}{(\partial_{ij}\psi)} \right)(\partial_ju_k)\partial_k\psi.
\end{equation}
\end{widetext}
Changing summation indices in order to factor the strains out, and using Stokes' theorem on the divergence term, we obtain that
\begin{widetext}
\begin{align}\label{eq:DF}
  \Delta \Free[\psi, \mathbf{u}] =-\int d^2 \mathbf{r} \left[ \pd{f}{(\partial_i\psi)}\partial_j\psi + \pd{f}{(\partial_{ik}\psi)}\partial_{jk}\psi - \left(\partial_k \pd{f}{(\partial_{ik}\psi)}\right)\partial_j\psi + \delta_{ij} f \right]\partial_iu_j
 - \int d S_i \pd{f}{(\partial_{ij}\psi)}(\partial_ju_k)\partial_k \psi,
\end{align}
\end{widetext}
where $d\mathbf{S}$ is the surface element vector on the boundary of the integration domain. Equation (\ref{eq:DF}) yields the elastic stress defined as the conjugate of the displacement gradient 
\begin{eqnarray}
  \sigma_{ij} &=& \frac{\delta\Delta \Free}{\delta(\partial_iu_j)} \nonumber\\
  &=& -\pd{f}{(\partial_i\psi)}\partial_j\psi 
- \pd{f}{(\partial_{ik}\psi)}\partial_{jk}\psi +\nonumber \\
&&+ \left(\partial_k \pd{f}{(\partial_{ik}\psi)}\right)\partial_j\psi
+ f\delta_{ij}.
\end{eqnarray}
For the specific free energy of PFC model from Eq.~(\ref{eq:F}), the stress field is given as 
\begin{align}\label{eq:sigmaij}
 \sigma_{ij} &= - \left[\mathcal{L}\psi\right]\partial_{ij}\psi + \left(\partial_i \mathcal{L}\psi\right)\partial_j\psi + f \delta_{ij}.\nonumber\\
  &= \left[\partial_i\mathcal{L}\psi\right]\partial_j\psi-\left[\mathcal{L}\psi\right]\partial_{ij}\psi + f \delta_{ij}, 
\end{align}
with $\mathcal{L} = 1 + \nabla^2$. Hence the elastic stress can be straightforwardly evaluated from the phase field $\psi$. Below we will show that this stress gives rise to the expected stress-strain relation in the linear elasticity regime.

The stress gives rise to a body force density $F_j = \partial_i \sigma_{ij}$ given by 
\begin{eqnarray}
  F_j &=& \nabla^2 \mathcal L\psi \partial_j \psi + \partial_i\mathcal L \psi \partial_{ij}\psi 
  - \partial_i \mathcal L\psi\partial_{ij}\psi \nonumber \\
  &&- \mathcal L \psi \partial_j\nabla^2\psi + \partial_{i} f\nonumber\\
  &=& \mathcal L^2\psi\partial_j \psi - \mathcal L \psi \mathcal L (\partial_j \psi) + \partial_{i} f.
\end{eqnarray}
If the medium is incompressible, the second term is the gradient of $-\frac 1 2 (\mathcal L \psi)^2$, and can be included into a pressure term $\partial_j p$ in the equation of conservation of momentum. Similarly, we can write 
\begin{equation}
F_j = \mu \partial_j \psi, 
\end{equation}
as the additional terms in the chemical potential $\mu = \frac{\delta \Free}{\delta \psi} = \mathcal L^2\psi + r\psi + \psi^3$ also lead to gradient terms.

If the medium is compressible, the additional contribution to the body force is given by
\begin{equation}
\partial_i \left( \delta_{ij} f \right) = \partial_j f = \mathcal L\psi \mathcal L (\partial_j \psi) + r\psi\partial_j\psi + \psi^3\partial_j \psi. 
\end{equation}
Hence, the body force in a compressible medium is the same as the incompressible case up to a gradient force and given simply as
\begin{equation}
F_j = \partial_i \sigma_{ij} = \left(\mathcal L^2\psi + r\psi + \psi^3\right)\partial_j\psi = \mu\partial_j \psi. 
\end{equation}
In short, the body force associated with small phase field distortions is simply given by $\mu\nabla\psi$.

For weak distortions, the stress can be written in terms of the amplitudes of Eq.~(\ref{eq:distortion}) in the one mode approximation. We first compute 
\begin{eqnarray}
\partial_i \psi &=& iA_0\sum_{|\mathbf q|=1} (q_i - q_k\partial_i u_k)\exp\left[i\mathbf q\cdot (\mathbf x - \mathbf u)\right], \nonumber\\
 \partial_{ij}\psi &=& -A_0\sum_{|\mathbf q|=1}\left( q_i q_j - q_j q_k\partial_i u_k - q_i q_k \partial_j u_k \right)\times\nonumber\\
 &&\times \exp\left[i\mathbf q\cdot(\mathbf x - \mathbf u)\right].
\end{eqnarray}
Therefore, it follows that 
\begin{align}
  \nabla^2\psi &= -A_0\sum_{|\mathbf q|=1} \left(1 - 2q_i q_k\partial_i u_k \right)\exp\left[ i\mathbf q\cdot(\mathbf x - \mathbf u)\right], \notag \\
 \mathcal{L}\psi &= \psi_0 + 2 A_0\partial_iu_k\sum_{|\mathbf q|=1} q_i q_k \exp\left[i\mathbf q\cdot(\mathbf x - \mathbf u)\right].
\end{align}
Changing summation indices and still assuming linear elasticity, we obtain
\begin{eqnarray}
  \partial_i\left(\mathcal{L}\psi\right) &=& 2iA_0\partial_lu_k\sum_{|\mathbf q|=1} q_l q_k q_i \exp\left[i\mathbf q\cdot (\mathbf x - \mathbf u)\right], \nonumber\\
  \left[\partial_i\left(\mathcal{L}\psi\right)\right]\partial_{j}\psi &=& -2A_0^2\partial_lu_k\sum_{\mathbf q, \mathbf q'}q_l q_k q_i q_j' \nonumber\\
  &&\times \exp\left[i(\mathbf q + \mathbf q')\cdot (\mathbf x - \mathbf u)\right].
\end{eqnarray}
Similarly, we compute
\begin{widetext}
\begin{align}
[\mathcal{L}\psi]\partial_{ij}\psi = &-\psi_0A_0\sum_{\mathbf q} \left(q_i q_j - q_i q_k \partial_ju_k - q_j q_k\partial_iu_k\right) \exp\left[i\mathbf q\cdot (\mathbf x - \mathbf u)\right] - 2A_0^2\partial_lu_k \sum_{\mathbf q, \mathbf q'}q_l q_k q_i' q_j' \exp\left[i(\mathbf q + \mathbf q')\cdot (\mathbf x - \mathbf u)\right].
\end{align}
\end{widetext}
Finally, by coarse graining over a unit cell of the lattice and taking the slowly-varying deformation gradients outside the integral, the single-$\mathbf{q}$ terms will 
vanish, while the $\exp[i(\mathbf{q} + \mathbf{q}')\cdot(\mathbf{x}  - \mathbf{u} )]$ factors integrate to $\delta_{\mathbf{q},-\mathbf{q}'}$. Therefore the averaged stress field from Eq.~(\ref{eq:sigmaij}) becomes 
\begin{align}
  \fv{\sigma_{ij}} &= 4A_0^2\partial_lu_k\sum_{|\mathbf q|=1}q_lq_kq_iq_j.
\end{align}
In tensorial form, using the three principal reciprocal lattice vectors $\mathbf q_n$ and including their negatives $-\mathbf q_n$ using a factor of $2$, this is equivalent to 
\begin{align}\label{eq:av_sigma}
  \fv{\bar{\bar\sigma}} &= 8A_0^2\sum_{n=1}^3 \mathbf q_n\mathbf q_n\mathbf q_n\cdot\nabla(\mathbf q_n\cdot \mathbf u).
\end{align}
Note that since coefficients of $\partial_lu_k$ are symmetric
under the interchange $l\leftrightarrow k$, we can also write the relation in terms of the symmetrized strain $u_{lk} = \frac 1 2 (\partial_lu_k + \partial_ku_l)$, as
\begin{equation}
\fv{\sigma_{ij}} = 8A_0^2u_{lk}\sum_{n=1}^3q^n_iq^n_jq^n_kq^n_l.
\label{eq:av_sigma_symm}
\end{equation}
Equation (\ref{eq:av_sigma_symm}) is a linear stress-strain relationship which only depends on the crystal reciprocal lattice vectors and the coarse grained or slowly varying amplitudes. For a hexagonal lattice,  inserting the reciprocal lattice vectors given in Eq.~(\ref{eq:reciprocal}) yields $C_{11} = C_{22} = 9A_0^2$, and $C_{12} = 3A_0^2$ and $C_{44} = 3A_0^2$ (cf., e.g., Ref.~\cite{elder2007phase}). This result can also be written in terms of Lam\'e coefficients as $\fv{\sigma_{ij}} = \lambda \delta_{ij}u_{kk} + 2\mu u_{ij}$ with $\lambda = \mu = 3A_0^2$, giving a Poisson's ratio of $\nu = \frac{\lambda}{2(\lambda+\mu)} = \frac 1 4$. This is different from the Poisson's ratio of $\frac 1 3$ obtained in Ref.~\cite{re:elder04}, as they use the plane stress condition, while we are assuming plane strain without loss of generality.

\section{Plastic flow and dislocation dynamics}

At the mesoscale level, the evolution of the phase field is driven by local relaxation of the free energy functional,
\begin{equation}
\frac{\partial\psi}{\partial t}= \nabla^2\frac{\delta \mathcal{F}}{\delta\psi}, 
\label{eq:pf_relaxation}
\end{equation}
where we have assumed a constant mobility coefficient (equal to unity in rescaled units). Equation (\ref{eq:pf_relaxation}) governs both conservation of mass and the evolution of crystal deformations. We will focus here on 2D systems, although a similar development can be applied in three dimensions.

There are no topological singularities in the phase field $\psi(\mathbf{r},t)$. However, under conditions in which the amplitude expansion of Eq.~(\ref{eq:one_mode_exp}) is valid (mean density $\psi_0$ and amplitudes $A_n$ that vary on length scales much larger than the wavelength of the reference pattern), topological defects can be identified from the location of the zeros of the complex amplitudes \cite{re:siggia81,re:shiwa86}. Evolution equations for $\psi_0$ and $A_n$ have been derived by several techniques, such as Renormalization Group methods~\cite{Goldenfeld2006} and multiple-scale analysis~\cite{Yeon10}. In the lowest derivative approximation that preserves the rotational invariance of the phase field model~\cite{Swinney94}, the resulting equations are given as~\cite{Yeon10} 
\begin{eqnarray}
 \frac{\partial \psi_0}{\partial t} =& \nabla^2\Big[(1+\nabla^2)\psi_0+\psi_0^3+6\psi_0\sum_n |A_n|^2\nonumber\\
&+6\left(\prod_nA_n+c.c.\right)\Big],\nonumber\\
 \frac{\partial A_n}{\partial t} =& -\mathcal{L}_n^2 A_n -(3\psi_0^2+r)A_n -6\psi_0 \prod_{m\ne n} A_m^*\nonumber\\
&-3 A_n\left(2\sum_m |A_m|^2-|A_n|^2\right),
\label{eq:Adot}
\end{eqnarray}
where $\mathcal{L}_n = \nabla^2+2i \mathbf q_n\cdot\nabla$. Spatial variations in $\psi_0$ in such a single component system need to be interpreted as a sign of the presence of vacancies, i.e., independent variations of $\psi_{0}$ and $\mathbf{u}$. 

The equations governing the evolution of the amplitudes are themselves variational, and can be written as \cite{Yeon10},
\begin{eqnarray}\label{eq:A_j}
 \frac{\partial \psi_0}{\partial t} &=& \nabla^2\frac{\delta \mathcal{F}_{CG}}{\delta\psi_0}\nonumber\\
 \frac{\partial A_n}{\partial t} &=& -\frac{\delta \mathcal{F}_{CG}}{\delta A_n^*}.
\end{eqnarray}
where $\mathcal{F}_{CG}\{\psi_0,A_n\}$ is the free energy, function of the amplitudes alone (a coarse grained free energy). Note however that all of these equations ignore higher amplitudes with $|\mathbf q| > 1$, so they are only valid at low quenches, $|r| \ll 1$.


\subsection{Transformation of field singularities to dislocation coordinates}
\label{sec:dislocs}
In order to make contact with the classical macroscopic description of plastic motion in terms of the velocity of a dislocation element under an imposed stress, we describe the transformation of variables that is required to relate the evolution of the phase field to the motion of its associated singularities. Assume a spatial distribution of discrete edge dislocations and define a Burger's vector density as $\mathbf{B}(\mathbf{r}) = \sum_\alpha \mathbf{b}_\alpha \delta(\mathbf{r} - \mathbf{r}_\alpha)$, where $\mathbf{r}_\alpha$ are the locations of the edge dislocation with Burger's vector $\mathbf{b}_\alpha$ in some element of volume. For each Burger's vector $\mathbf b_\alpha$ we define the three integers $s_n^\alpha = \frac{1}{2\pi}(\mathbf q_n \cdot \mathbf b_\alpha)$, which satisfy the relation $\sum_{n=1}^3 s_n^\alpha = \frac{1}{2\pi}\mathbf b_\alpha \cdot \sum_{n=1}^3\mathbf q_n = 0$.

An edge dislocation at $\mathbf r_\alpha$ corresponds to a deformation field $\mathbf u(\mathbf r)$ with $\oint d\mathbf u = \mathbf b_\alpha$ around a contour containing only $\mathbf r_\alpha$. This deformation field is associated with a phase factor in the complex amplitudes, given by $A_n(\mathbf r) = |A_n| e^{-i\mathbf q_n \cdot \mathbf u + i\phi}$, with $\phi(\mathbf r)$ smooth inside the contour. The phase circulation of the amplitude around the same contour can then be found as
\begin{align}
\oint d(\operatorname{arg}A_n) &= -q^n_j\oint \partial_k u_j dr_k + \oint \partial_k \phi dr_k \notag \\
&= -q^n_j b_j^\alpha = -2\pi s_n^\alpha,
\end{align}
using that $\phi$ has no circulation, being smooth inside the contour. Thus the amplitude $A_n$ has a vortex with winding number $-s_n^\alpha$ at $\mathbf r = \mathbf r_\alpha$. This gives the transformation of delta functions\cite{re:halperin81,Mazenko97,mazenko2001defect,Angheluta12}
\begin{align}
D_n\delta(A_n) &= 
-\sum_\alpha s_n^\alpha\delta(\mathbf{r} - \mathbf{r}_\alpha) \notag\\
&= -\frac{1}{2\pi}\sum_\alpha (\mathbf q_n \cdot \mathbf b_\alpha)\delta(\mathbf r - \mathbf r_\alpha),
\label{eq:deltatransform}
\end{align}
for a given amplitude $A_n$, where 
\begin{equation}
D_n = \operatorname{Im}\left( \partial_xA_n^*\partial_yA_n \right) = \frac{1}{2i}\epsilon_{ij}\partial_iA_n^*\partial_jA_n, 
\end{equation}
is the Jacobian of the transformation from complex amplitudes $A_n$ to vortex coordinates $\mathbf{r}_\alpha$. Multiplying the above expression with a reciprocal vector $\mathbf{q}_n$ and summing over $n$, we find the dislocation density as
\begin{equation}
\label{eq:b}
\mathbf{B}(\mathbf{r}) = -\frac{4\pi}{3}\sum_{n=1}^3 \mathbf{q}_nD_n\delta(A_n), 
\end{equation}
making use of the fact that $\sum_{n=1}^3 q^n_iq^n_j = \frac 3 2 \delta_{ij}$ (see appendix \ref{sec:currents} for why we use reciprocal lattice vectors in this expansion rather than real space lattice vectors).

In order to obtain the equation governing the motion of the Burgers vector density, we use that the determinant fields $D_n$ have conserved currents given by~\cite{Angheluta12}
\begin{equation}
\label{eq:J}
J^{(n)}_k = \frac{1}{2i} \epsilon_{kl}\left( \dot A_n \partial_lA_n^* - \dot A_n^* \partial_lA_n \right) = \epsilon_{kl}\operatorname{Im}\left( \dot A_n\partial_lA_n^* \right), 
\end{equation}
so that $\partial_t D_n = - \partial_k J^{(n)}_k$, as can be verified by insertion. The amplitude evolution at the vortex location $\dot A_n$ can be found from an amplitude expansion of $\dot \psi$, such as Eq.~(\ref{eq:Adot}).

We also have a similar continuity equation for the delta functions,
\begin{equation}
D_n \pd{}{t}\delta(A_n) = -J^{(n)}_i \partial_i \delta(A_n).
\end{equation}
This can be proved by differentiating through the delta functions and inserting for $ D_n$ and $ J^{(n)}_i$, giving
\begin{align}
- J_i^{(n)}\partial_i \delta(A_n) &= \frac{i}{2}\epsilon_{ij}\left(\dot A_n\partial_jA_n^* - \dot A_n^* \partial_jA_n\right)
\partial_iA_n\delta'(A_n) \notag\\
&= \frac i 2 \epsilon_{ij} \partial_jA_n^*\partial_iA_n\dot A_n\delta'(A_n) \notag\\ 
&=  D_n \pd{}{t}\delta(A_n).
\end{align}
Hence, differentiating the dislocation density with time, we find the Burger's vector current 
\begin{align}
  \pd{B_i}{t} &= -\frac{4\pi}{3}\sum_{n=1}^3q^n_i \left( \pd{D_n}{t}\delta(A_n) + D_n \pd{}{t}\delta(A_n) \right) \notag\\
  &= \frac{4\pi}{3}\sum_{n=1}^3 q^n_i\left( \partial_j J_j^{(n)}\delta(A_n) + J_j^{(n)}\partial_j\delta(A_n) \right) \notag \\
  &= \partial_j\left(\frac{4\pi}{3}\sum_{n=1}^3 q^n_i J_j^{(n)}\delta(A_n) \right) = -\partial_j \mathcal{J}_{ij}.
\end{align}
Whenever $D_n = 0$ we have $\delta(A_n) = 0$, otherwise we can transform back to physical coordinates using Eq.~(\ref{eq:deltatransform}),
\begin{align}
\mathcal{J}_{ij} &= - \frac{4\pi}{3}\sum_{n=1}^3q_i^nJ_j^{(n)}\delta(A_n) \notag\\
&= \frac{2}{3}\sum_{n=1}^3 q_i^n J_j^{(n)}
\sum_\alpha \frac{\mathbf{q}_n\cdot\mathbf{b}_\alpha}{ D_n} \delta(\mathbf{r} - \mathbf{r}_\alpha),
\label{eq:Jij}
\end{align}
ignoring diverging terms.
On the other hand, if the dislocations are moving with velocity $\mathbf{v}_\alpha$, we have
%
%
\begin{equation}
\mathcal{J}_{ij} = \sum_\alpha b_i^\alpha v_j^\alpha\delta(\mathbf r - \mathbf r_\alpha). \label{eq:current_v}
\end{equation} 
This gives an equation for the velocity of the dislocation indexed by $\alpha$, 
\begin{equation}
b_i^\alpha v_j^\alpha = \frac 2 3 b_k^\alpha\sum_{n=1}^3 q_i^nq_k^n \frac{J_j^{(n)}}{ D_n},
\end{equation}
which is solved by contracting with the Burger's vector to give
\begin{align}
v_j^\alpha &= \frac 2 3 \sum_{n=1}^3\frac{(\mathbf q_n\cdot \mathbf b_\alpha)^2}{|\mathbf b_\alpha|^2}\frac{ J_j^{(n)}}{ D_n} \notag \\
&= \frac{1}{S_\alpha^2}\sum_{n=1}^3(s_n^\alpha)^2 \frac{ J_j^{(n)}}{D_n}, \label{eq:bv}
\end{align}
where we set $S_\alpha^2 = \sum_{n=1}^3 (s_n^\alpha)^2$ and used that
$|\mathbf b_\alpha|^2 = \frac 8 3\pi^2S_\alpha^2$. This is a general result and the central relation between the velocity of a point singularity and the equation governing the evolution of the phase field amplitudes. We apply this expression below to obtain the velocity response of a single edge dislocation under an applied strain.

\subsection{Dislocation motion}

At a dislocation core, assumed at $\mathbf r = 0$, the amplitude $A_n$ will vanish
as long as $2\pi s_n = \mathbf q_n \cdot \mathbf b \neq 0$. Since $s_1 + s_2 + s_3 = 0$, any dislocation must give rise to vortices in at least two of the three amplitudes, and so these two amplitudes vanish. This means that the amplitude evolution equation (\ref{eq:Adot}) at the dislocation position reduces to 
\begin{equation}
\dot A_n(\mathbf r=0) \approx -\mathcal{L}_n^2 A_n\Big|_{\mathbf r =0}, 
\end{equation}
so that the amplitudes decouple and we can study the vortex motion independently for each amplitude.
We assume that the dislocation is stationary in the absence of external stress, which means that
$\mathcal L_n^2 A_n\Big|_{\mathbf r=0} = 0$.

The crystal is then perturbed by an affine deformation $\mathbf u$ from an external load, so that the amplitudes transform as $A_n \mapsto \tilde A_n = A_{n} e^{-i\mathbf q_n \cdot \mathbf u}$. This will cause the dislocation to move in response to the applied force, i.e. 
\begin{equation}
\partial_t \tilde A_n = -\mathcal L_n^2 \tilde A_n \neq 0,
\end{equation}
and our aim is to compute how the resulting dislocation motion depends on the deformation. Let us focus on one deformed amplitude $\tilde A_n$ and call it $\tilde A = Ae^{-i\mathbf q\cdot \mathbf u}$, with its associated wave vector $\mathbf q$. 

In the limit of small deformations, we have
\begin{align}
  \partial_i \tilde A &= \left( \partial_i A - iA q_k\partial_i u_k\right)e^{-i\mathbf q \cdot \mathbf u} \nonumber\\
  \partial_{ij} \tilde A &=\Big(\partial_{ij} A - i\partial_i A q_k\partial_j u_k -i\partial_j A q_k \partial_i u_k\nonumber\\
	&-iA q_k \partial_{ij} u_k - A q_k q_l \partial_i u_k\partial_j u_l \Big)e^{-i\mathbf q\cdot \mathbf u}\nonumber\\
	&\approx \left(\partial_{ij} A - i\partial_i A q_k\partial_j u_k - i\partial_j A q_k \partial_i u_k\right)e^{-i\mathbf q\cdot \mathbf u}.
\end{align}
Continuing in this manner and using that $A$ is a stationary vortex solution, we then have that 
\begin{eqnarray} 
\partial_t \tilde A &=& -\mathcal L^2 \tilde A = 
4iq_j [(\partial_i + iq_i)\mathcal LA] \partial_iu_je^{i\mathbf q \cdot \mathbf u}.
\end{eqnarray}
If $s = \pm 1$, $\mathcal L^2 A = 0$ is solved by the isotropic vortex solution $A \propto x - isy$, for which $\mathcal L A = 2iq_k\partial_k A$, and $\partial_i \mathcal L A = 0$. Hence $\partial_t \tilde{A}$ is simplified to 
\begin{equation}
\partial_t \tilde A =
-8iq_i q_j q_k\partial_k A \partial_iu_je^{i\mathbf q \cdot \mathbf u}.
\end{equation}
\begin{figure*}
  \centering
   \includegraphics[width=0.8\textwidth]{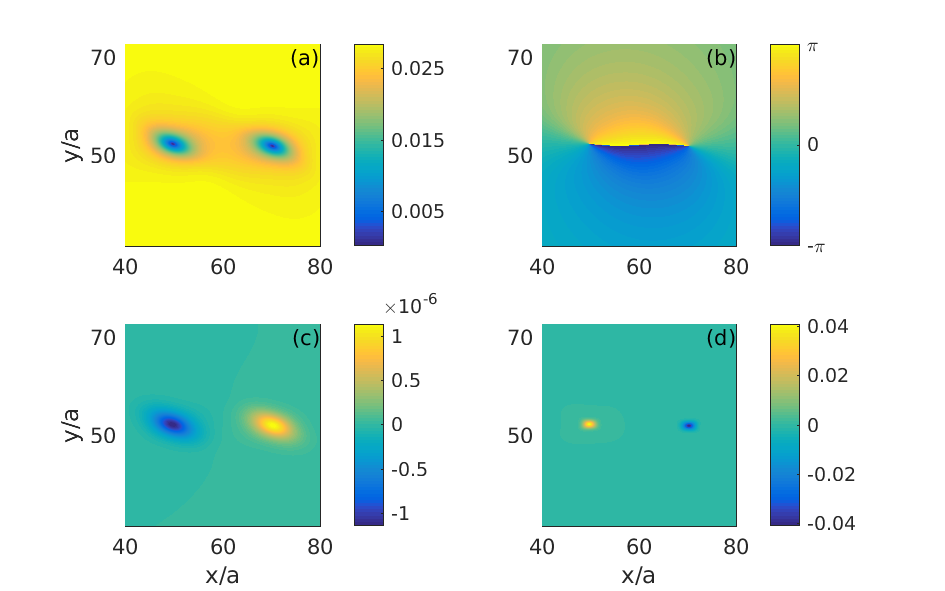}
 \caption{(a) and (b): Magnitude and phase of the $A_2$ amplitude, showing the initial vortices corresponding to the initial dislocations. (c): The $D_2$ field showing the sign of the vortex charge. (d): The resulting $B_x$ dislocation density in the $x$ direction, with $w = A_0/5$. $x$ and $y$ is given in units of the lattice constant $a = \frac{4\pi}{\sqrt 3 q_0}$.} 
  \label{fig:1}
\end{figure*}
%
We now calculate the defect current corresponding to a non-conserved order parameter (the complex amplitude) from Eq.~(\ref{eq:J}) as
\begin{eqnarray}
 J_i &=& \epsilon_{ij}\Im(\partial_t\tilde A\partial_j \tilde A^*)\nonumber\\
&=& -8\epsilon_{ij}q_kq_l q_m \partial_k u_l\Im\left(i\partial_m A\partial_j A^*\right)
\end{eqnarray}
for the corresponding defect density $\rho(\mathbf r,t) = \delta(A)D = q\delta(\mathbf r)$. Since the defect density is unchanged under the smooth deformation, the Jacobi determinant at the dislocation position is unchanged,
\begin{equation}
D = \frac{1}{2i}\epsilon_{ij}\partial_i \tilde A^*\partial_j \tilde A = \frac{1}{2i}\epsilon_{ij}\partial_i A^*\partial_j A.
\end{equation}
The isotropic vortex $A\propto x - isy$ satisfies 
\begin{equation}
i\partial_{i} A = -\frac 1 s \epsilon_{ij}\partial_j A,
\end{equation}
so that 
\begin{equation}
 J_i = \frac 8 s\epsilon_{ij}\epsilon_{mo}q_k q_l q_m \partial_k u_l  \Im\left(\partial_o A\partial_j A^*\right). 
\end{equation}
We can compute that $\Im(\partial_o A\partial_j A^*)=\epsilon_{jo}D$, which means that
\begin{equation}
 J_i = \frac 8 s\epsilon_{ij} q_j q_k q_l \partial_k u_l D.
\end{equation}
Thus, for a simple dislocation with all $|s_n| \le 1$, we find that the vortex velocity from Eq.~(\ref{eq:bv}) is
\begin{eqnarray}
v_i &=& \frac{8\epsilon_{ij}}{S^2}\sum_{n=1}^3s_nq^n_jq^n_kq^n_l\partial_ku_l \notag \\
&=& \frac{4b_m}{\pi S^2}\epsilon_{ij}\sum_{n=1}^3 q^n_m q^n_j q^n_kq^n_l\partial_ku_l \nonumber \\ 
& = & \frac{1}{4\pi A_0^2}\epsilon_{ij}\fv{\sigma_{jk}} b_k,
\label{eq:dislocation_velocity}
\end{eqnarray}
by using that the deformation gradient is related by the stress-strain relation from Eq.~(\ref{eq:av_sigma}) to the {\em elastic} stress. Explicit in the derivation is the exclusion of singular strains or any local variations in the amplitudes. Thus, we obtain an expression for the dislocation velocity determined by the Peach-Koehler force as in classical dislocation models,  e.g.,~\cite{groma2010statistical}, but with the same mobility coefficient for both climb and glide motion. This particular result of an isotropic mobility follows as a consequence of the one mode amplitude expansion for the phase field employed. This is a valid approximation at low quenches ($|r| \ll 1$). We have checked numerically that the defect analysis presented works well also at deep quenches (finite $r$), where one cannot use the one-mode amplitude expansion, as discussed in the next section. 

\begin{figure*}
  \centering
   \includegraphics[width=0.8\textwidth]{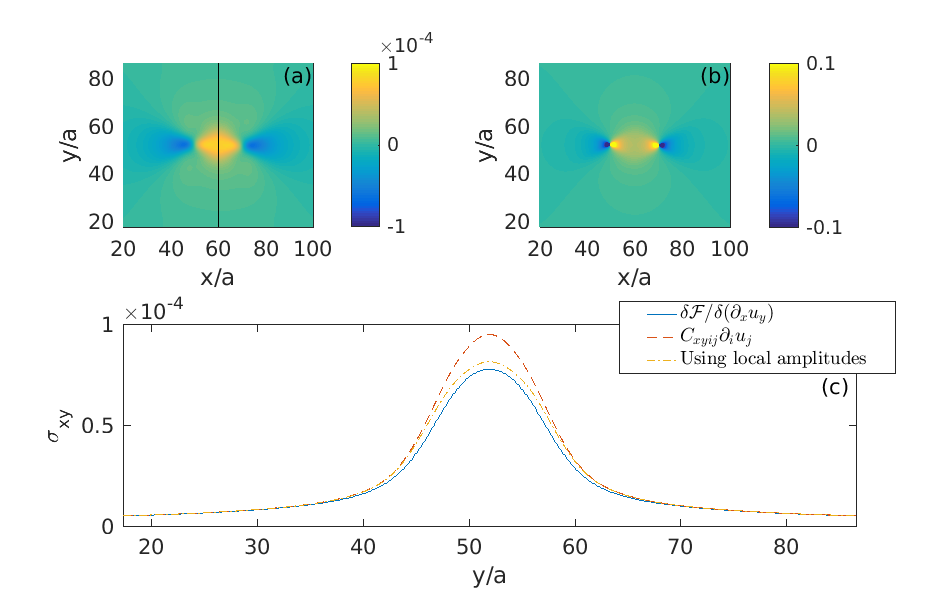}
 \caption{(a): Map of the stress field $\fv{\sigma_{xy}}$, as computed directly from the formula in eq.~(\ref{eq:sigmaij}), with a Gaussian average. (b): Map of the strain field $\partial_y u_x$, computed from the amplitudes by eq.~(\ref{eq:strain}). (c): Comparison of the stress computed along the indicated line in three different ways: Using the direct expression for the stress (solid line), using the stress-strain relation with the equilibrium amplitude (dashed line), and using the stress-strain relation with the average local amplitude $\frac 1 3(|A_1|^2 + |A_2|^2 + |A_3|^2)$. All the expressions agree where the amplitude is in equilibrium. Deviations occur where the amplitude deviates from equilibrium, but can be partly corrected for by using the local value.}
  \label{fig:2}
\end{figure*}

\section{Numerical results}
We test our analytical predictions by directly simulating a simple hexagonal crystal containing a dislocation dipole. We use two parameter sets for probing low and deep quenches regimes, i.e. $r = -0.01$ and $\psi_0 = -0.04$ (low quench) and $r = -0.8$ and $\psi_0 = -0.43$ (deep quench). 

The initial state is prepared by setting $\psi(\mathbf r) = \psi_0 + \sum_n A_n e^{i\mathbf q_n \cdot \mathbf r} + c.c.$, where the amplitudes contain vortices with the appropriate charges for each dislocation, e.g. $A_n = A_0\exp\left[-\sum_\alpha i s_n^\alpha \theta(\mathbf r - \mathbf r_\alpha)\right]$. We then evolve Eq.~(\ref{eq:pf_relaxation}) using an exponential time differencing method~\cite{cox2002exponential}, and track the motion of dislocations as topological defects.

The amplitudes of a phase field are computed by performing a local amplitude decomposition, which corresponds to averaging $\psi e^{-i\mathbf q \cdot \mathbf r}$ over a region roughly corresponding to a unit cell~\cite{re:guo15}. For numerical stability we use a convolution with a Gaussian of width $a = 2\pi/\sqrt 3$ instead of hard limits to the averaging region. This convolution is most efficiently evaluated in Fourier space, using the expression
\begin{equation}
A_n(\mathbf r) = e^{-i\mathbf{q_n\cdot\mathbf r}} F^{-1}\left[ e^{-\frac 8 3\pi^2(\mathbf k-\mathbf q_n)^2} F[\psi] \right],\label{eq:ampdec}
\end{equation}
where $F$ and $F^{-1}$ denote the Fourier and inverse Fourier transforms, respectively.
Figure~\ref{fig:1} shows the magnitude and phase of the complex amplitude $A_2$ for the initial dislocation dipole after a short period of relaxation (panels a-b). 
\begin{figure}
  \centering
  \includegraphics[width=0.43\textwidth]{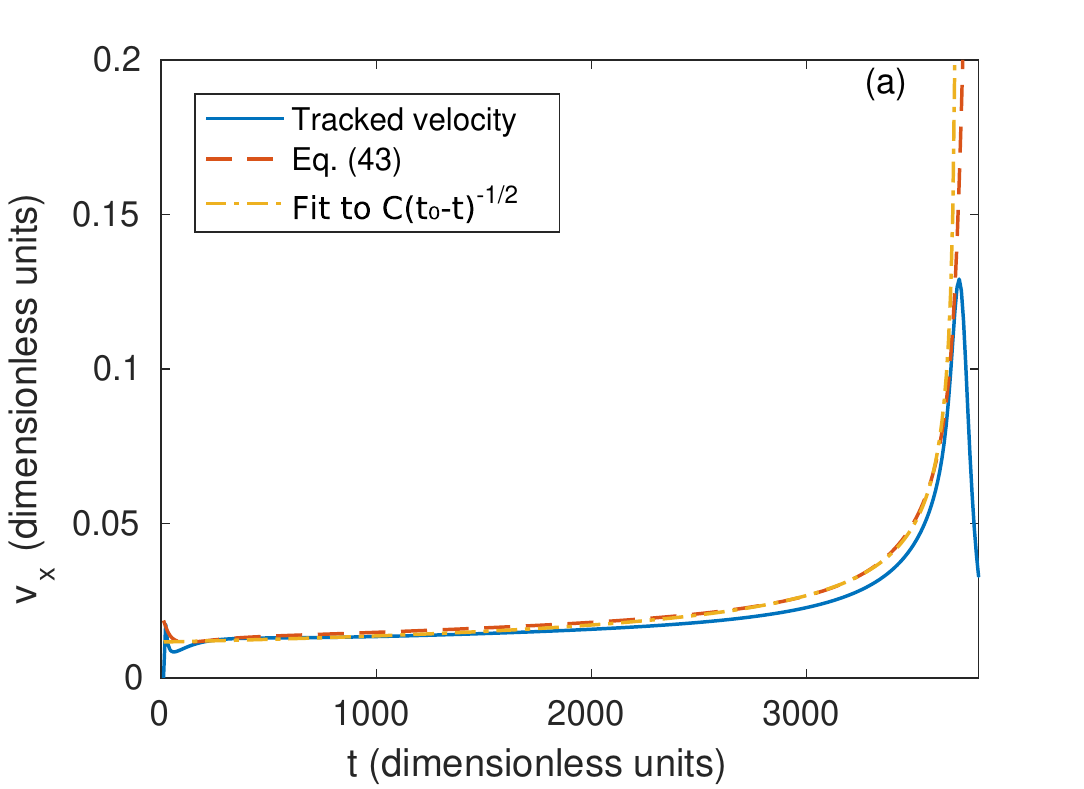}
  \includegraphics[width=0.43\textwidth]{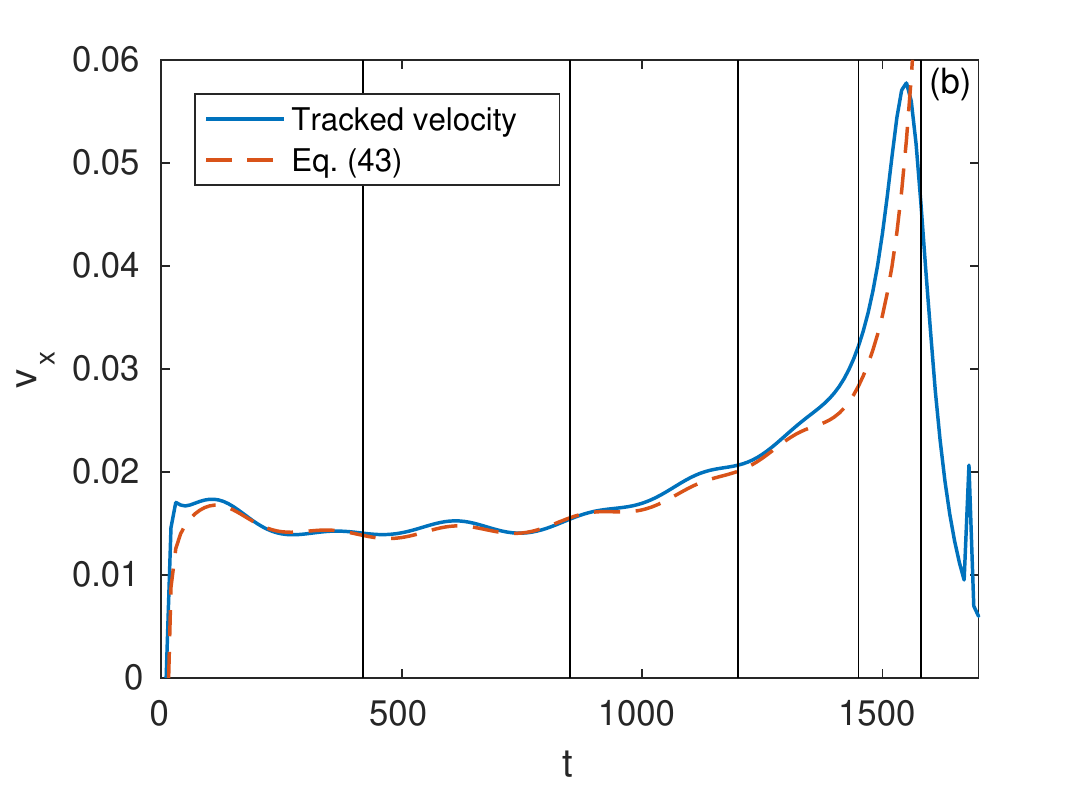}
 \caption{The dislocation velocity as a function of time until the annihilation time for low quenches (panel a) versus deep quenches (panel b), given in the dimensionless units of Eq.~(\ref{eq:pf_relaxation}). Vertical lines indicate points in time where the dislocation has traveled a distance $a$ from its initial point.}
  \label{fig:3}
\end{figure}

From the amplitudes we can calculate a Gaussian approximation to the $\delta(A_n)$ function, by 
\begin{equation}
\delta(A_n) = \frac{1}{2\pi w^2}e^{-\frac{|A_n|^2}{2w^2}},
\end{equation}
where smaller $w$'s give sharper delta functions. Along with the $D_n$ fields obtained by numerically differentiating the amplitudes (Fig.~\ref{fig:1}, panel c), we obtain approximations to the Burger's vector density from Eq.~(\ref{eq:b}), shown in Fig.~\ref{fig:1}, panel (d). 
Thresholding these fields allows us to track the positions of the Burger's vectors, which also gives an estimate of the dislocation velocity.

Using that $A_n = |A_n|e^{-i\mathbf q_n \cdot \mathbf u}$, we find that 
\begin{equation}
\operatorname{Im}\frac{\partial_jA_n}{A_n} = -q_k^n\partial_ju_k,
\end{equation}
which can be inverted to find 
\begin{equation}
\partial_ju_k = -\frac 2 3 \sum_n q^n_k \operatorname{Im}\frac{\partial_jA_n}{A_n},
\label{eq:strain}
\end{equation}
giving numerical values for the strains. The shear strain obtained by this analysis is plotted in Fig.~\ref{fig:2} together with the shear stress derived from the phase field free energy in Eq.~(\ref{eq:sigmaij}). Additionally, we plot the shear stress as a function of y along a particular line and verify that it is well reproduced using the strain fields and the stress-strain relation, Eq.~(\ref{eq:av_sigma_symm}).      

The amplitude evolution $\dot A_n$ can be found in two ways: Either by using Eq.~(\ref{eq:Adot}) directly, which is valid for low quenches, or by employing Eq.~(\ref{eq:ampdec}) to find the amplitudes of $\dot \psi$. Both ways allow us to compute the currents $J_j^{(n)}$, which are then used in Eq.~(\ref{eq:bv}) to extract the dislocation velocity. The result is shown in Fig.~\ref{fig:3} for the dislocation velocity at low quenches (panel (b) ) versus deep quenches (panel (b)) as a function of time. At low quenches, the dislocations move towards each other according to Peach-Koehler force until they annihilate. It is expected that given that the elastic shear stress decays as $1/r$, $r$ being the distance between dislocations, the velocity will increase with time as $v\sim (t_0-t)^{-1/2}$ with $t_0$ representing the annihilation event. This is also shown in panel (a). However, at deeper quenches we notice that the dislocation velocity varies non-monotonically and shows a stick and slip like behavior with periodicity related to the lattice constant $a$ (panel (b)), consistent with previous numerical simulations from Ref.~\cite{Berry2006}. These are lattice effects on the motion of the amplitudes when $r$ is not small, the phase field analog of Peierls stresses \cite{re:boyer02b}.

\section{Conclusions and discussion}

We have introduced a coarse graining procedure of a phase field model of a crystalline phase that reveals the topological charge of an isolated dislocation from the regular (non singular) phase field itself. This is accomplished through consideration of the slowly varying amplitudes or envelopes of the phase field in the vicinity of the defect. The amplitudes allow the computation of local elastic stresses at the defect, as well as the derivation of an exact relation between the velocity of the point defect and the kinetic equations governing the evolution of the amplitudes. The combination of both results allows the derivation of the Peach-Koehler force on the defect, as well as an explicit derivation of the defect mobility. A parallel coarse graining procedure of a numerically determined phase field has been introduced, and used to verify the analytic results for the case of the motion of a dislocation dipole in a two dimensional hexagonal lattice.

Phase field crystal models of the type discussed in this paper lack a dependence on lattice deformation as an independent variable. However, we have shown explicitly that it is possible to calculate the {\em elastic} stress directly from the phase field free energy by considering its variation with respect to a suitably chose phase field distortion. The stress thus derived is consistent with linear elasticity and leads to known expressions for the elastic constants for the phase field crystal. Furthermore, the phase field description can also describe defected configurations. While the phase field remains non singular no matter how large the local distortion of the reference configuration is, the location of any isolated singularities can be accomplished through the determination of the zeros of a slowly varying (on the scale of the periodicity of the field) complex amplitude or envelope of the phase field. Such a coarse graining is essential to define singular fields from the regular phase field. On this slow scale, we have then derived the Peach-Koehler force on a topological defect. As expected, this force depends only on the slowly varying stress (distortion), and not from any other fast variations of the phase field near the defect.

Our results also clarify the relationship between dissipative relaxation of the phase field and plastic motion. Equation (\ref{eq:dislocation_velocity}) relates the velocity of a dislocation with its Burgers vector and the {\em slowly} varying stress $\langle \sigma_{ij} \rangle$. Such a relation follows directly from the equation governing the relaxation of the phase field, Eq. (\ref{eq:pf_relaxation}), in the range of $r \ll 1$ in which it can be described by an amplitude equation. This equation also gives an explicit expression for the dislocation mobility which depends on the specific functional form of the free energy considered. Of course, any fast variations of the phase field near defects are still described and very much included in Eq. (\ref{eq:pf_relaxation}). Short scale effects such as dislocation creation and annihilation, and any nonlinearities of both elastic and plastic origin evolve according to the dissipative evolution of the phase field. The free energy involved in this dissipative evolution also serves to define a Burgers vector scale, and topological charge conservation over large length scales. 


\acknowledgments{This research has been supported by a startup grant from the University of Oslo, and by the National Science Foundation under contract DMS 1435372}  

\appendix
\section{Fully determining the dislocation current}
\label{sec:currents}
Equation (\ref{eq:b}) gives an expression for the dislocation density in terms of the three reciprocal lattice vectors $\mathbf q_n$. Since the Burger's vector is a vector in the real lattice, it would seem more natural to express the dislocation density in terms of the two real space lattice vectors $\mathbf a_n$, where $\mathbf q_n \cdot \mathbf a_m = 2\pi \delta_{mn}$ (for $n,m = 1,2$). Indeed, using that $\sum_{n=1}^2 a^n_i q^n_j = 2\pi \delta_{ij}$, we find the alternative expression
\begin{equation}
\mathbf B(\mathbf r) = -\sum_{n=1}^2 \mathbf a_n D_n \delta(A_n),
\end{equation}
which of course is equal to from Eq.~(\ref{eq:b}).
Going through the same derivation as in section \ref{sec:dislocs} leads to a Burger's vector current 
\begin{equation}
\mathcal J_{ij} = -\sum_{n=1}^2 a^n_i J_j^{(n)}\delta(A_n),
\end{equation}
however this current does not agree with the current in Eq.~(\ref{eq:Jij}).

The missing point is that the conservation equation for the field $D_n \delta(A_n)$,
\begin{equation}
\partial_t [D_n\delta(A_n)] + \partial_i [J_i^{(n)}\delta(A_n)] = 0,
\end{equation}
only determines its current $I_j^{(n)}$ up to an unknown divergence-free vector field $K_j^{(n)}$, i.e. 
\begin{equation}
I_i^{(n)} = J_i^{(n)}\delta(A_n) + K_i^{(n)},
\end{equation}
where $\partial_i K_i^{(n)} = 0$. To determine this residual current, we observe that 
\begin{equation}
\sum_{n=1}^3 D_n \delta(A_n) = -\frac{1}{2\pi}\sum_\alpha b^\alpha_i\delta(\mathbf r - \mathbf r_\alpha) \sum_{n=1}^3q^n_i = 0,
\end{equation}
due to the resonance condition $\sum_n \mathbf q_n = 0$. Hence it is natural to require that the current of this field vanishes identically, 
\begin{equation}
\sum_{n=1}^3 I_i^{(n)} = \sum_{n=1}^3 J_i^{(n)}\delta(A_n) + \sum_{n=1}^3 K_i^{(n)} = 0.
\end{equation}
This condition is fulfilled by setting $K_i^{(n)} = -\frac 1 3 \sum_{m=1}^3 J_i^{(m)}\delta(A_m)$, which has vanishing divergence. With this choice, the dislocation current in Eq.~({\ref{eq:Jij}}) is modified to 
\begin{equation}
\mathcal J_{ij} = -\frac{4\pi}{3}\sum_{n=1}^3 q_i^n J_j^{(n)}\delta(A_n) + \frac{4\pi}{9} \sum_{n=1}^3 q_i^n \sum_{m=1}^3 J_j^{(m)}\delta(A_m),
\end{equation}
where the second term vanishes due to resonance. Hence the additional fields $K_i^{(n)}$ give no contribution when we express $\mathbf{B}(\mathbf r)$ in terms of the three reciprocal lattice vectors. On the other hand, if we used real lattice vectors $\mathbf a_n$ instead, the extra term would not vanish.

\bibliographystyle{apsrev4-1}
\bibliography{ref}

\end{document}